\documentclass[preprint,3p,times,showpacs,preprintnumbers,amsmath,amssymb,superscriptaddress,floatfix]{elsarticle}

\usepackage{tabularx}
\usepackage{graphicx}
\usepackage{color}
\usepackage{xcolor}
\usepackage{dcolumn}
\usepackage{bm}
\usepackage[latin1]{inputenc}
\newcommand*{\rom}[1]{\expandafter\@slowromancap\romannumeral #1@}
\usepackage{epstopdf}
\usepackage{amsmath}
\usepackage{multirow}
\usepackage[normalem]{ulem}
\useunder{\uline}{\ul}{}
\usepackage{caption}
\usepackage{hyperref}
\usepackage{lineno}
\usepackage{arydshln}

\biboptions{numbers,sort&compress}

\journal{arXiv}

\begin{document}

\begin{frontmatter}

\title{GIS-AHP Multi-Decision-Criteria-Analysis for the Optimal Location of Solar Energy Plants at Indonesia}

\author[label1]{H. S. Ruiz}
\ead{dr.harold.ruiz@leicester.ac.uk}

\author[label2]{A. Sunarso}
\ead{sunarso@dosen.polnep.ac.id}

\author[label1,label2]{K. Ibrahim-Bathis}
\ead{ib132@leicester.ac.uk}

\author[label1,label2]{S. A. Murti}
\ead{sm1015@leicester.ac.uk}

\author[label1,label2]{I. Budiarto}
\ead{ib131@leicester.ac.uk}

\address[label1]{Aerospace and Computational Engineering Group \& Space Park Leicester, School of Engineering, University of Leicester, University Road, Leicester, LE1 7RH, United Kingdom}

\address[label2]{Department of Mechanical Engineering, Politeknik Negeri Pontianak, Jalan Ahmad Yani, Pontianak, ZIP 8124, Indonesia}

\begin{abstract}

A reliable tool for site-suitability assessment of solar power plants capable to account for the sustainable development and protection of cultural and biodiversity conservation areas is proposed. We present a novel Analytic Hierarchy Process (AHP) based approach for the Multi-Decision Criteria Analysis (MDCA) of SSI satellite retrieved data and local information sources, it within a GIS platform tailored to fit the needs of energy stakeholders at Indonesia, simultaneously ensuring the conservation of legally protected areas. This imposes significant challenges due to the wide diversification of cultural, natural, and ecological protected areas that need to be considered, in landmarks that demand for high resolution imaging of surface solar irradiance (SSI) within $\pm 4^{\circ}$ of the equator. To overcome these challenges, a GIS spatial weighted overlay analysis  for criteria layers such as climatology, topography, electrical grid, and road infrastructure has been performed, it based on the technical and economic feasibility for solar plants deployment within three approximation schemes focused on their proximity to the existing (i) power network, (ii) road infrastructure, and (iii) community settlements. Here, we focused on the West Kalimantan Province of Borneo Island (WKP), it mainly due to its possibility of onshore inter-connectivity and energy share with Malaysia and Brunei, and the high national and international importance that brings forward the protection of the biodiversity of Borneo. It has been found that the optimal location of PV plants can be reduced to just $0.03\%$ ($46.60~km^{2}$) and $0.07\%$ ($108.58~{km}^{2}$) of WKP, in what we report as the best-suitable conditions out of the $33.05\%$ exploitable area found after the exclusion of conservation areas. This corresponds to an estimated annual generation of $8.91-20.96~{TWh/year}$ for conventional PV farms, resulting in a daily generation capacity of approximately $43.65~{MW/km^{2}}$, where about $8.4~{km^2}$ of PV deployment will be sufficient to meet the national energy targets.

\end{abstract}


\end{frontmatter}



\section{Introduction}\label{Section_1}

Being the third largest island in the world and with a privileged location over the equator, the island of Borneo is considered as one of the richest landmasses for renewable power generation, currently attracting substantial attention from governments and investors onto the search of adequate routes for planning and exploitation of their solar-energy resources. Politically divided among three countries, Brunei, Malaysia, and Indonesia, with the latter occupying up to $73\%$ of Borneo's territory, Borneo's existing map for the expansion of their power generation and transmission sectors faces a considerably large number of societal, political and technical challenges, some of these caused by (i) their highly-decentralized power networks, (ii) their already overburden urban grids by the rapid growth in population and industrialization sectors, (iii) their large dependence on oil and coal based energy generators for off-grid rural substations, and (iv) the large but vital constraints on the use of land and right-of-way imposed by the protection of their natural resources and conservation legal frameworks. However, it is the expectation of the Association of Southeast Asian Nations (ASEAN), that by 2030 the Indonesian territory of Kalimantan together with the two Malaysian states of Sarawak and Sabah, will have become a major energy resource center within an interconnected power network~\cite{ADB2014}, benefiting of a strong intake of renewable energy resources~\cite{Irena2017} for meeting the UN Sustainable Development goals. However, whilst the state-owned energy provider Sarawak Energy is taking the lead on this milestone, according to the figures disclosed by the Ministry of Utilities of Sarawak last December 2019 in the Borneo's Sustainability \& Renewable Energy Forum~\cite{Saref2019a,Saref2019b}, the case of Indonesian's Borneo results more complex to analyze, it due to the more diverse set of legal constraints implied, and its high dependence on fossil-derived energy sources. In fact, the Indonesian situation is much critical than the Malaysian one, as just during the last five years Indonesia has imported about 200 million barrels of oil for the energy sector, with an approximate cost of 150 trillion IDR (~$\$11.5$ billion) yearly~\cite{PWC2018}, resulting in a severe burden on the national budget whilst threatening the national energy security with compulsory power outages for under-developed communities~\cite{IESR2019}. 

This situation, added to the lack of reliable data for the deployment of other renewables, has led to the national electricity company owned by the Indonesian government, PLN~\cite{PLN}, which controls nearly $100\%$ of the power distribution and generation sectors across the country, to issue of a new procurement plan in 2018 which overall shaves further additions of energy stakes as a result of weaker-than-expected electricity demand growth, but sticking with coal- fired generation as the main technology choice where existing geothermal plants cannot be considered as the primary option. Still, in the most recent Indonesia's 2019-2028 Electricity Procurement Plan (RUPTL) issued  on February 20, 2019 by the Ministry of Energy and Mineral Resources in conjunction with PLN and the Department of Population and Civil Registration~\cite{PLN_RUPTL2019}, RUPTL prioritizes the use of renewable energy sources as the basis for PLN development business plans. For that purpose, also hydro-power and photovoltaic energy are being prioritized, with the latter being subject to the identification of optimal locations for the deployment of solar farms, meeting a minimum target of $1~{GW}$ solar power plant development per year until 2028. This is certainly a conservative figure given the average insolation measure of $4.80~{kWh/m^{2}/day}$~\cite{Veldhuis2015}, which can render to a bare estimation of about $500~{GW}$ of solar power potential~\cite{IESR2019}, but simultaneously denotes the main challenge that Indonesia is currently facing on the investment procurement for solar energy, which is no other than the identification of optimal locations where protected and conservation areas are not a impediment for the expansion of PLN grid~\cite{Hamdi2019,Irena2019}. 

\begin{figure}
\begin{center}
{\includegraphics[width=0.65\textwidth]{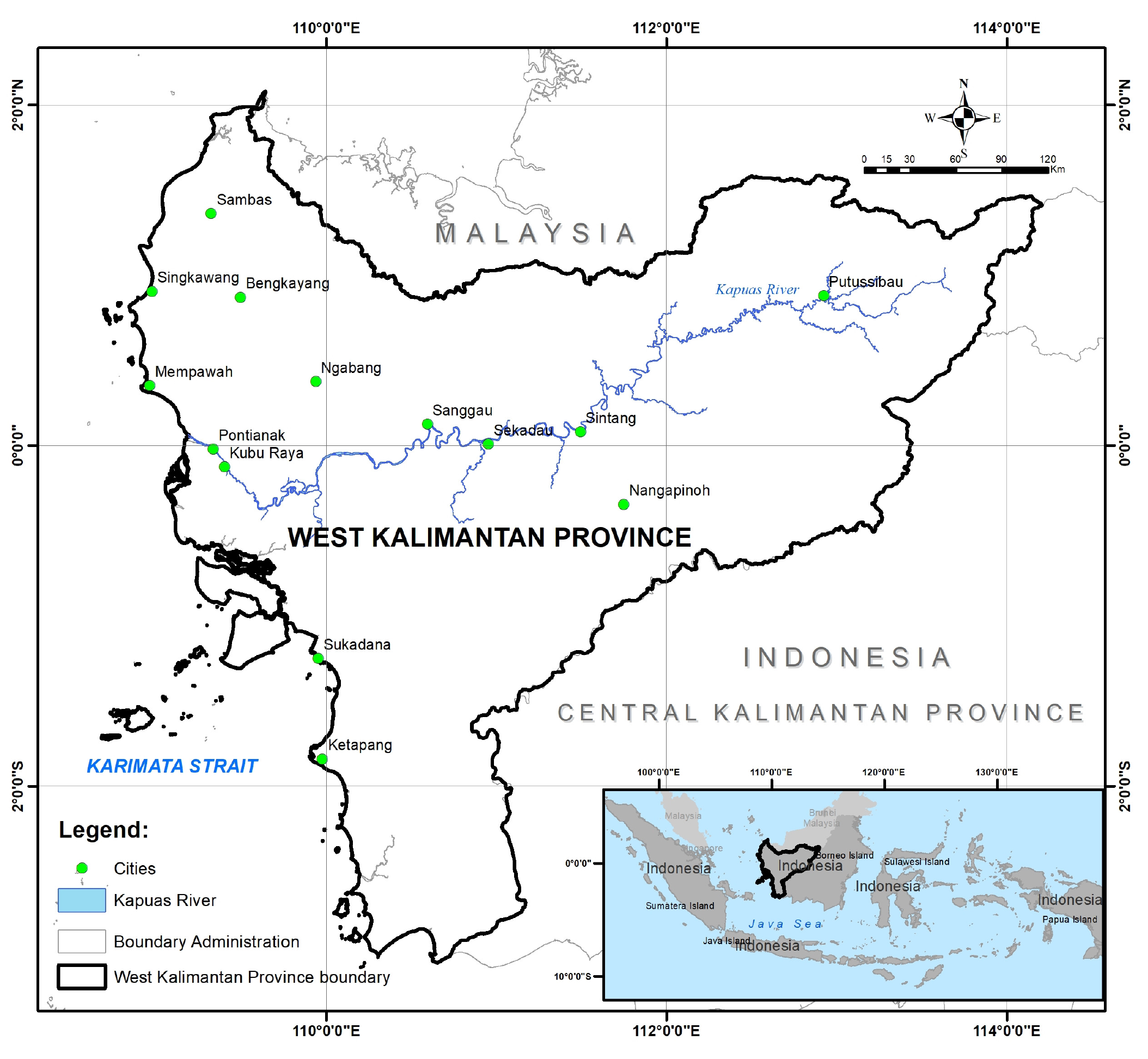}}
\caption{Location map of the West Kalimantan Province (WKP) at Indonesia's Borneo.}
\label{Figure_1}
\end{center}
\end{figure}

In fact, if Indonesia wants to meet its energy mix target by 2030, the power sector has already targeted the development of solar PV plants with capacity of $6.5~{GW}$ in 2025 and $14.2~{GW}$ in 2030~\cite{RUEN2017}. To achieve this ambitious goal, solar PV plants are expected to be developed across 34 provinces in Indonesia, from which the Kalimantan region is expected to have a share of $1.081~{GW}$ in 2025, with $366.4~{MW}$ assigned to the West Kalimantan Province (WKP), $232.1~{MW}$ for East Kalimantan, $221.0~{MW}$ for Central Kalimantan, $160.0~{MW}$ for South Kalimantan, and $101.7~{MW}$ for North Kalimantan. However, the country already struggles with a limited grid-connection and poor transportation network, causing higher equipment delivery costs for the deployment of solar power farms, particularly in rural and remote areas where the energy demand is still high and the electrification ratio is low~\cite{PWC2018,Hamdi2019,Irena2019,Rose2016}. Additionally, the lack of an appropriate regulatory support system for the energy tariffs between different regencies, the unstable political and administrative coordination between local governments and stakeholders, and virtually non-existing tools for site-planing and land constraints identification, discourage investors on pursuing developing and planning permits as well as land acquisition bills for solar energy projects. Nevertheless, given the proximity of the existing power network at the West Kalimantan Province (WKP) of Indonesia (see Fig.~\ref{Figure_1}), and the largest of the Malaysian states (Sarawak), the energy prospects of the WKP for renewable sources as solar energy present a singular interest in the Borneo Region, although the planning of these facilities demand of a careful revision of the climatology and topography  conditions of the province, as well as the influence of legal and natural constraints imposed by the Indonesian government and business stakeholders. Moreover, it is worth mentioning that between the different renewable sources available at Indonesia~\cite{Hasan2012}, geothermal exploration in Borneo is not an option, as this is legally classified as a mining activity that can endanger the sustainability of protected and conservation areas~\cite{Kumar2016}, but whose privileged location over the equator gives it certain privilege for the consideration of solar farms.

WKP is the second largest and most populated province at Borneo,  which according to the latest report facilitated to us by PLN~\cite{PLN_Stat2019}, it currently has less than $0.01\%$ of solar installed capacity $(0.18~{MW})$, within a total WKP generation capacity of $200.99~{MW}$, from which more than $60\%$ of it $(123.03~{MW})$ comes from Diesel powered generators. Therefore, designing an adequate planning tool for the identification of optimal locations of solar power plants at WKP has been identified as a priority for the Indonesian government, and in general of Borneo Island, being this the central subject of this paper which reports on the main results of the 2019-2020 British Council Newton Fund Institutional Links project ``Solarboost''~\cite{solarboost}. For this purpose, we have counted with the active participation of major Indonesian stakeholders such as PLN~\cite{PLN}, the Ministry of Energy and Mineral Resources (ESDM~\cite{ESDM}), the Regional Development Planning Agency (BAPPEDA~\cite{BAPPEDA}), the Ministry of Environment and Forestry (MENLHK~\cite{MENLHK}), the Ministry of Agriculture (PERTANIAN~\cite{PERTANIAN}), the Ministry of Public Works and Public Housing (PUPR~\cite{PUPR}), and Indonesian Meteorology Climatology and Geophysics Council (BMKG~\cite{BMKG}), which all have helped us into the conceptual design of our Geographic Information System (GIS), aided by a Multi-Decision Criteria Analysis model (MDCA) and an integrated Analytical Hierarchy Process (AHP) for policy making decisions, which ultimately helps to eliminate the communication and administrative barriers that are currently impeding the successful deployment of the solar sector across the region. Our system allows to identify optimal locations for solar power plants aimed to potential energy investors, based on a weighted overlay analysis of satellite retrieved high-resolution surface solar irradiation data, aided by relevant topographical, land use, and infrastructure data layers provided by our stakeholders at WKP. 

Consequently, in this paper we present a comprehensive study about the main barriers that, within a data-oriented approach, must be incorporated into a GIS-AHP algorithm for the deployment of solar energy plants in landscapes with robust challenges as the ones encountered in the WKP at the Borneo Island.  In processing all these spatial data, including vulnerability zones, protected and conservation land areas for analysing the solar energy feasibility in WKP, in Section~\ref{Section_2} we introduce a Multi-Decision Criteria Analysis (MDCA) model integrated within an AHP algorithm which renders the analytical techniques for achieving higher effectiveness of GIS decision systems with multiple layers of information (see Fig.~\ref{Figure_2}), in similar way to the recent works of H.~Z.~Al-Garni et. al. on Saudi Arabia~\cite{AlGarni2017} and H. E. Colak on the province of Malatya in Turkey~\cite{Colak2020}. Then, in Section~\ref{Section_3} the results obtained from three different MDCA schemes are described, and the comparisons between their performances and subsequent implications are demonstrated, such that an optimal scheme and consequently an optimal area for the installations of solar power plants at WKP is disclosed. Finally, the main conclusions of this paper are presented in Section~\ref{Section_4}.


\section{GIS-AHP-MDCA Model Framework}~\label{Section_2}

\begin{figure*}[!]
\begin{center}
{\includegraphics[width=0.6\textwidth]{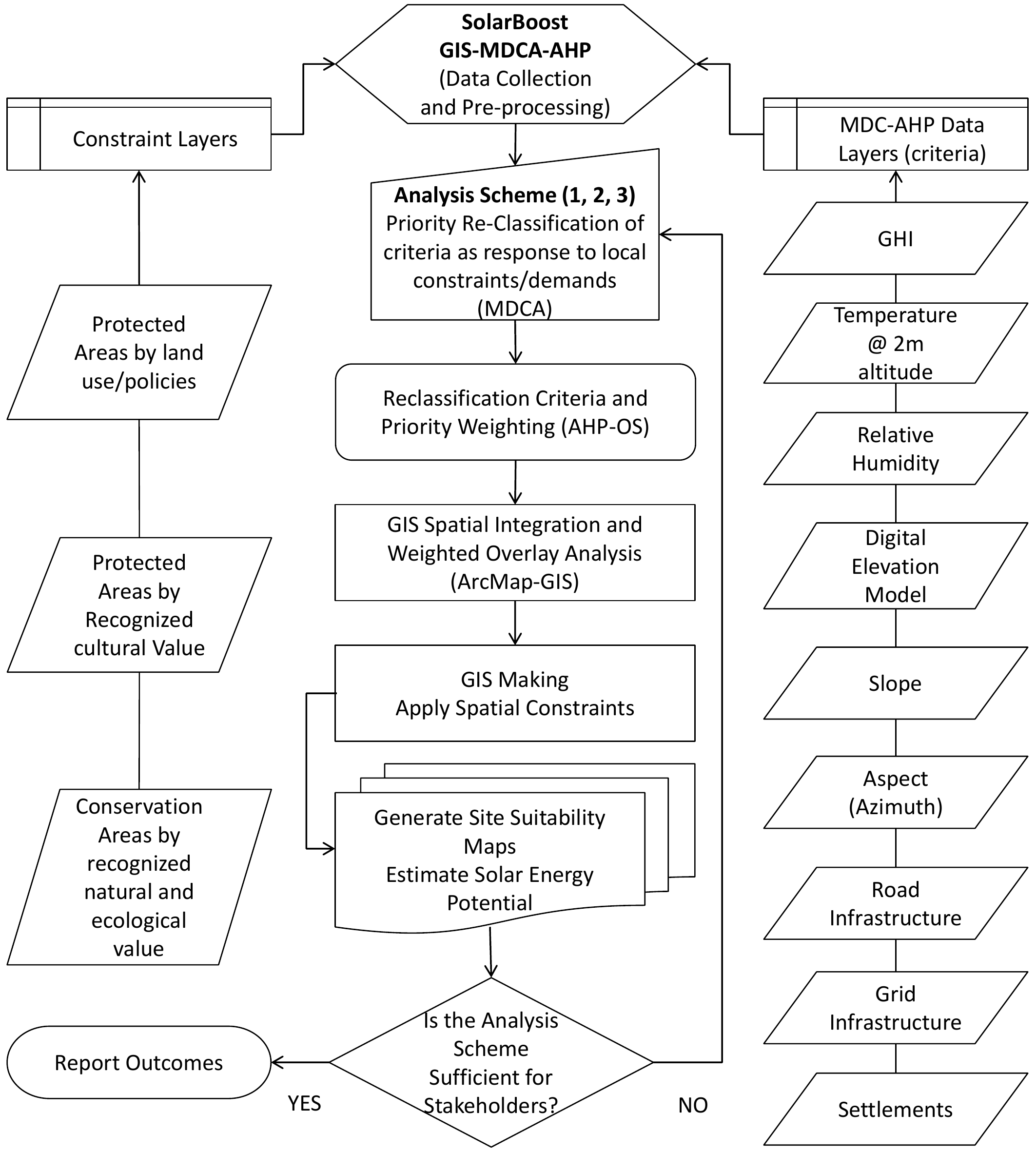}}
\caption{Working flowchart of the GIS-AHP-MDCA platform developed for this study, which has been called SolarBoost to make reference to the funder's project identifier.}
\label{Figure_2}
\end{center}
\end{figure*}

GIS-based Multi Criteria Decision Analysis (GIS-MDCA) techniques aided by AHP models for generating maps of potential areas for solar power plants development, in varying climate and topographic conditions, have been the subject of intense research during the last few years~\cite{AlGarni2017,Colak2020,Shorabeh2019,Giamalaki2019,Majumdar2019,Doorga2019,Yousefi2018,
Doljak2017,Zoghi2017,Alami2016,Noorollahi2016,Kucuksari2014,Sanchez2013,Uyan2013}. However, a common factor between all this literature is that the availability of feasible sites for large scale solar farms deployment, results highly determined by the different criteria that could have been selected in the study factors, and their corresponding weighting which are assigned within the AHP algorithm. Therefore, although there is no strict protocol in assigning weighting factors between the assessed criteria when these are not clearly linked by a known physical variable, like it is the case, for instance, for the Global Horizontal Irradiation factor (GHI) and its relation with the proximity of a solar farm with relevant local infrastructures (power network, roads, settlements, etc), then the assigning of relative weighting factors mostly depends on the decision of the researchers after their consultancy with relevant stakeholders and policymakers. To illustrate this, in Table~\ref{Table_1} we present a brief comparison of some of the most relevant literature where different AHP weightings have been assigned to the same criteria, but based on the understanding of the authors of the local conditions and further constraints of the scrutinized area, as well as the state-of-the-art literature at the time of the publications (see \cite{AlGarni2017,Colak2020,Shorabeh2019,Giamalaki2019,Majumdar2019,Doorga2019,Yousefi2018,
Doljak2017,Zoghi2017,Alami2016,Noorollahi2016,Sanchez2013,Uyan2013} and references therein). Within this, we can see that regarding to the development of solar farms, the GHI is generally considered as the evaluation criteria with the greatest weighting factor, in particular when the area of a full country is being analyzed. However, when a better insight of the local climate, topology, energy-related policies, and the stakeholders interests is known, a larger weighting can be given to other particular criteria such as, for instance, the dust storms in Isfahan-Iran~\cite{Zoghi2017}, or the proximity to the existing power grid in Cartagena-Spain~\cite{Sanchez2013}. Therefore, if we want to provide a reasonable estimation of the criteria of interest and their corresponding weighting factors in a GIS-AHP algorithm for Borneo island, bearing in mind that ultimately this must be conceived as a platform aimed for the promptly use of solar energy investors and local policy-makers, we argue that the best and most sensible strategy is to focus the study first at a province level (e.g., at WKP), where decisions could be made promptly without encountering inter-regional administrative and political barriers. 

In consequence, the present section of this study focuses on the formulated methodology for finding the optimal location or best suitable area for large scale solar power plants $(>5~{MW})$, with special emphasis on the retrieved data from local information sources at WKP in Indonesia's Borneo. This enables the further consideration of vulnerability zones such as Borneo's conservation and land protected areas, as well as other information layers with relevant climate, topology, settlement, and proximity to infrastructure (roads and power grid) data, which are all together analyzed with respect to their spatial interrelationship with the local GHI map and its derived solar energy potential. In brief, a GIS-AHP MDCA method is employed to assign the rank and priority factors for the information layers aforementioned, where for the sake of simplicity and accessibility to any stakeholder, we have taken advantage of the recently introduced AHP Online Software (AHP-OS) by K. Goepel~\cite{Goepel2018}, and then, a comprehensive spatial weighted overlay integration analysis is performed in ArcMap-GIS software (v10.6.1) following the overall methodology depicted in Figure~\ref{Figure_2}. 


\begin{table*}[!]%

\caption{\label{Table_1} AHP weighting factors for different evaluation criteria extracted from the literature.}


\begin{center}

\begin{tabular}{ccccccc} \hline \hline

\textbf{Evaluation} & Saudi & Mauritius & Serbia & Isfahan & Iran & Cartagena\\
\textbf{Criteria} & Arabia~\cite{AlGarni2017} & \cite{Doorga2019} & \cite{Doljak2017} & Iran~\cite{Zoghi2017} & \cite{Noorollahi2016} & Spain~\cite{Sanchez2013} \\
\hline \\
Solar radiation (GHI) & 0.322 & 0.401 & 0.305 & 0.25 & 0.275 & 0.238 \\ 
Sunshine radiation (daylight hours) & $^{\dag}$ & 0.131 & 0.184 & 0.19 & $^{\dag}$ & $^{\dag}$\\ 
Air temperature & 0.243 & 0.033 & 0.111 & $\dag$ & 0.071 & 0.048\\
Relative Humidity & $\dag$ & 0.016 & 0.048	 & 0.043 & 0.041 & $\dag$\\
Elevation & $\dag$ & 0.021 & $\dag$ & 0.059 & 0.081 & $\dag$\\
Slope & 0.163 & 0.194 & 0.153 & 0.042 & 0.08 & 0.112\\
Land Aspects/Use & 0.108 & 0.046 & 0.077 & 0.066 & 0.07 & 0.116\\
Vegetation & $\dag$ & $\dag$ & 0.122 & $\dag$  & $\dag$  & $\dag$\\
Clouds/Snow/Rain/Dust conditions & $\dag$  & $\dag$  & $\dag$ &  0.254	0.101 & $\dag$\\
Proximity to Power Grid & 0.085 & 0.093 & $\dag$ & 0.05 & 0.112 & 0.415\\
Proximity to Road Infrastructure & 0.046 & 0.065 & $\dag$ & 0.032 & 0.088 & 0.043\\
Proximity to Settlements & 0.032 & $\dag$ & $\dag$ & 0.014 & 0.081 & 0.028\\
\hline \hline
\end{tabular}
\\
$^{\dag}$ \footnotesize{This criteria has either, not been considered within the cited study or it is not of relevance for the study location.}
\end{center}
\end{table*}
%

Firstly, concerning to the collection of data for the GIS-AHP MDCA platform, we are specifically interested in the West Kalimantan Province which extents between $2\textdegree 08^{\prime}$~N and $3\textdegree 02^{\prime}$~S, and between $108\textdegree 33^{\prime}$~E and $104\textdegree 10^{\prime}$~E, covering a total area of $146,807~{km}^{2}$, with 14 major cities/settlements distributed across the province as show in Fig.~\ref{Figure_1}. With the equator line crossing across this region the climate conditions of WKP are fairly uniform and ideal for the deployment of solar farms as reported by BMKG~\cite{BMKG}, with a daily average of sunlight of $12$~hr and $7$~min (sunrise $\sim 05:50$, sunset $\sim 17:57$), an annual average relative humidity of more than $90\%$, and a daily mean temperature at 2 meter height which varies between $25.9\textdegree$C and $28.4\textdegree$C across the entire year. Therefore, daylight is not a criteria to be considered in this study as it will be irrelevant into a MDCA, but temperature and humidity are weighting criteria that still need to be considered, this mainly due to the fact that the efficiency of photo-voltaic (PV) cells and solar mirrors, both strongly depend on the variance of these two factors. However, as the annual mean temperature and relative humidity variance across WKP is almost negligible $(<2\%)$, they can be defined as a low priority criteria. Additionally, the WKP is benefited by a low slope topology which is ideal for minimizing aggregated costs of land leveling and flattering, both commonly required for the deployment of solar farms~\cite{Sabo2016,Sabo2017}, what makes also of this factor a low weighting criteria. Here, it is worth emphasizing that this study is not limited to PV farms but instead considers the whole solar energy prognostics, where other means of solar energy capture can be conceived such as solar thermal collectors. This explains why our study is focused on relatively large solar power plants $(>5~{MW})$ which are aimed to be connected to the main electrical grid, instead of off-grid approaches for small communities where roof PV-cells might be sufficient. 


\begin{table*}%

\caption{\label{Table_2} GIS layers defining local constraints.}


\begin{center}

\begin{tabular}{ccc} \hline \hline

\textbf{Data Layer} & \textbf{Source} & \textbf{Constraint}\\
\hline
Settlement Location & \cite{IGIA2014s} & Settlement Proximity (Sp) \\
Power Grid & \cite{PLN,INEC2019} & Grid Proximity (Gp)\\
Road Network$^{\dag}$ & \cite{IMPW2015} & Road Proximity (Rp)\\
Forestry Area & \cite{MENLHK,IMF2014} & Protected area by law enforcement No. 733/2014 and no. 45/2004\\
Wildlife or Endangered habitat & \cite{BKSDA2011} & Protected area by law enforcement No. 308/MENLHK/2019\\
Peatland & \cite{IMF2014} & Protected area by law enforcement No. 8599/MENLHK/2018\\
Cultural/Community Forest & \cite{IMF2014} & Protected area by law enforcement No.21/MENLHK/2019\\
Water bodies & \cite{IGIA2014r} & Protected area by law enforcement No. 38/2011\\
Rice field & \cite{IMA2012} & Protected area by law enforcement No.1/Perda Kalbar/2018\\
\hline \hline
\end{tabular}
\\
$^{\dag}$ \footnotesize{Data retrieved in paper format.}
\end{center}
\end{table*}
%

Likewise, in what concerns to the intensity of Global Horizontal Solar Irradiation, having chosen WKP as case of study is not an arbitrary decision as by analysing the data recorded by Solargis and The World Bank Group~\cite{WBG2017,WBG2019}, the highest potential of total global horizontal irradiation (GHI) per year is recorded in the city of Ketapang at the SE part of WKP (see fig.~\ref{Figure_1}) with $1742~{kWh/m2/year}$. Nevertheless, this figure is subject to a maximum uncertainty of $\pm 8\%$ due to the high humidity level of the region, which adds another reason for which  the humidity criteria needs to be accounted even if it has a low priority factor. Actually, the GHI, topology, and climatology data layers which can be collected from online public domains, all together serve as the base framework for any GIS-AHP MDCA model regardless of the location (see Table~\ref{Table_1}), but it is the local data provided by relevant stakeholders what can give an actual meaning and impact for choosing one or another MDCA-AHP scheme. In this sense, the different data layers collected from local stakeholders at WKP are summarized within Tables~\ref{Table_2} $\&$ \ref{Table_3}, where the full set of GIS criteria layers, sources of information, and mapping constraints are disclosed. 

\subsection{GIS Constrained Data Layers}~\label{Section_2.1}

\begin{figure}[t]
\begin{center}
{\includegraphics[width=0.65\textwidth]{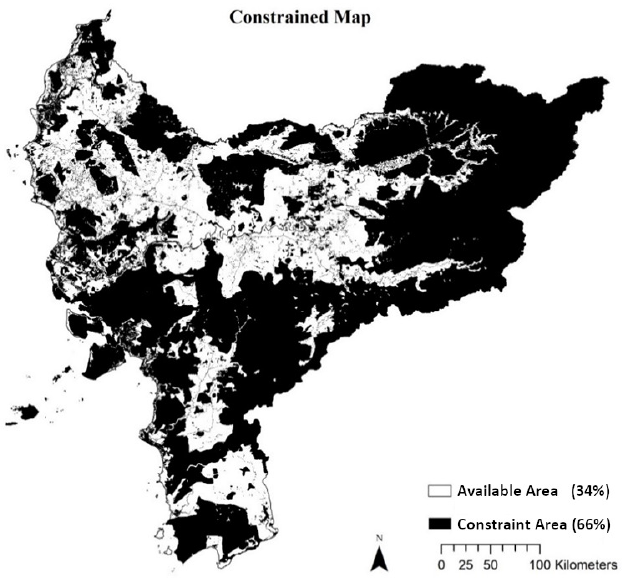}}
\caption{Map showing the total calculated constraint region at WKP (in black), defining prohibited areas for solar farms deployment as reported in Table~\ref{Table_2}. Source Maps can be consulted at our created Web-Gis within the SolarBoost project~\cite{solarboost} or by request to the author of correspondence.}
\label{Figure_3}
\end{center}
\end{figure}

In a hierarchy process, all data layers mentioned in Table~\ref{Table_2} are initially classified as secondary maps as these have to be preprocessed or digitalized for adequate rendering into the GIS platform. Thus, any non-spatial data collected from the different sources are converted into the spatial data and assigned UTM projection system, where north and south UTM grid zones must be considered for area calculations given that the WKP lies partially in both sides of the equator. Some of these data layers are considered as constraint layers as these involve protected land factors dictated by national and regional laws, where energy investments and related works might be undermined or simply prohibited by the sole existence of these factors. In particular, we have found that about $66\%$ of the total land area of WKP (See Fig.~\ref{Figure_3}) is subjected to legal constraints where any kind of developmental activities or inversion on these lands, different to the one originally intended, is heavily penalized by law enforcement due to their potential threat to the environmental and ecological equilibrium at local, national, or even global scale, as defined by the Indonesian government.


\begin{table*}[t]%

\caption{\label{Table_3} GIS-AHP-MDCA classification criteria layers subdivided into three major factors: (i) Climatology: GHI, Temperature, and Relative Humidity, (ii) Topography: Elevation, Slope, and Aspect, and (iii) Proximity factor: Power Grid, Road infrastructure, and Major Settlements. }


\begin{center}

\begin{tabular}{ccccc} \hline \hline

\textbf{AHP Factor} & \textbf{Source} & \textbf{Range} & \textbf{Class Range $\Delta$} & \textbf{AHP Grading 1-9 Class Limits}\\
\hline
GHI & \cite{WBG2019} & $2.6-5.04$ & $0.28$ & Grade n = 1 to 9 class limits @ \\
$[kWh/m^{2}]$ & & & & GHI$=2.6+n*\Delta$ \\ 
\hdashline
 & & & & Grade 1 class @ $T>27.8$,\\
Temperature $(T)$~[\textdegree C] & \cite{WBG2019} & $17.1-27.8$ & 1 & Grade n = 2 to 8 class @ $27-(n-1)*\Delta$ \\
 & & & & Grade 9 class @ $27-(n-1)*\Delta$ \\
\hdashline  
Relative  & &  & & Grade 1 class @ $H > 91$,\\
Humidity $(H)$  & \cite{NASA} & $81.99-91.58$ & $1$ & Grade n = 2 to 8 class @ $91-(n-1)*\Delta$ \\  
$[ \% ]$ & & & & Grade 9 class @ $H < 84$ \\  
\hdashline 
\hdashline
Elevation & \cite{CGIAR2020} & $<90$ & $10$ & Grade n = 1 to 9 class limits @ \\
$(DEM)~[m]$ & & & & $DEM=90-n*\Delta$ \\ 
\hdashline
Slope & \cite{CGIAR2020} & $<9$ & $1$ & Grade n = 1 to 9 class limits @ \\
$(S)~[ \% ]$ & & & & $S=n*\Delta$ \\ 
\hdashline
 & & & & Grade 9 class @ ,\\
 &  &  & N , NE & N(00-22.50) \& S(157.50-202.50) \\
Aspect &  & WKP & E , SE & Grade 5 class @\\
Azimuth & \cite{CGIAR2020} & UTM & S , SW & NE(22.50-67.50) SE(112.50-157.50) \\
$(A_{z})$ & & Zones & W, NW & SW(202.50-247.50) NW(292.50-337.50) \\
 & & & & Grade 1 class @ \\
 & & & & E(67.50-112.50) \& W(247.50-292.50) \\
\hdashline
\hdashline
Road Proximity & \cite{IMPW2015} & $0.1 \leq R_{p} \leq 10$ & $1.1$ & Grade n = 1 to 9 class limits @ \\
$(R_{p})~[km]$ & & & & $R_{p}=10-n*\Delta$ \\ 
\hdashline
Power Grid & \cite{INEC2019} & $0.1 \leq G_{p} \leq 10$ & $1.1$ & Grade n = 1 to 9 class limits @ \\
Proximity $(G_{p})~[km]$ & & & & $G_{p}=10-n*\Delta$ \\ 
\hdashline
Major Settlements & \cite{IGIA2014s} & $0.5 \leq S_{p} \leq 10$ & $1.055$ & Grade n = 1 to 9 class limits @ \\
Proximity $(S_{p})~[km]$ & & & & $R_{p}=10-n*\Delta$ \\ 
\hline \hline
\end{tabular}
\\
\end{center}
\end{table*}
%

\begin{figure}[!]
\begin{center}
{\includegraphics[width=0.97\textwidth]{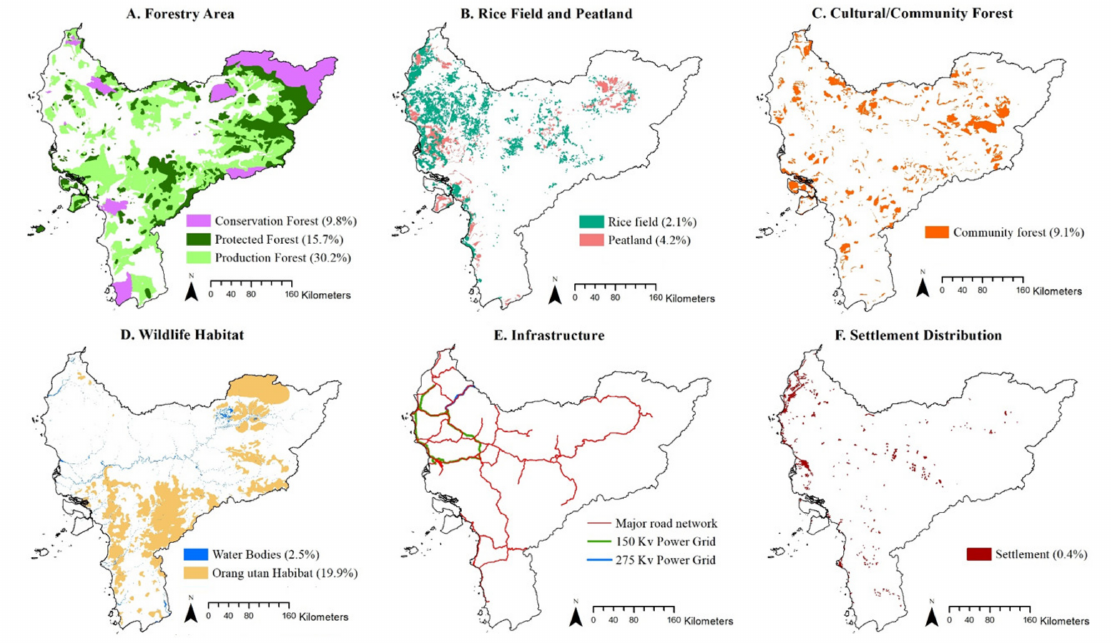}}
\caption{Thematic maps used as constraint layers for site exclusions within the MDCA-AHP suitability analysis for the deployment of large scale solar PV plants in WKP. Source Maps can be consulted at our created Web-Gis within the SolarBoost project~\cite{solarboost} or by request to the author of correspondence.}
\label{Figure_4}
\end{center}
\end{figure}

The constraint layers which are particularly defined by the enforcement of national or regional laws, and therefore demand liaison with relevant stakeholders~\cite{PLN,ESDM,BAPPEDA,MENLHK,PERTANIAN,PUPR,BMKG}, or at least a comprehensive knowledge of relevant governmental laws and the local language, can be classified between different subgroups for easy GIS visualization (see Table~\ref{Table_2}). Between these, we can find: (i) Forestry Areas (Fig.~\ref{Figure_4}A), which include factors such as production forestry ($30.2\%$ of WKP), protected forests $(15.7\%)$ and conservation forests $(9.8\%)$, (ii) Production and conservation weatlands (Fig.~\ref{Figure_4}B), including rice fields $(2.1\%)$ and peatlands $(4.2\%)$, (iii) Protected cultural and community (Fig.~\ref{Figure_4}C) forests $(9.1\%)$, (iv) Wildlife habitats (Fig.~\ref{Figure_4}D) which account for water bodies $(2.5\%)$ and protected Orang-Utans habitats $(19.9\%)$, (v) Relevant infrastructure (Fig.~\ref{Figure_4}E) such as, the major road network, and power transmission grid and, finally, (vi) the settlements which comprehend $0.4\%$ of the WKP area (Fig.~\ref{Figure_4}F).  Thus, by overlapping all these data-layers, we have found that at least $66\%$ of the total area must be extracted or constrained from the general GIS within MDCA-AHP (Fig.~\ref{Figure_3}), in order to determine the optimal location for the installation or foreseeable deployment of solar power plants at WKP. However, it is important to mention that although other constraint layers belonging to relevant productivity areas at Indonesia could be included, such as oil palm and mining areas, these areas are not being included in this study as the current energy plans of the country aim for the reduction of its dependence on these sources~\cite{RUEN2017,RUPTL,PLN_RUPTL2019}, and furthermore, these are not considered as protection or conservation areas as per the law. 

\subsection{GIS Methodology for assessing AHP Weighted Data Layers in the MDCA}~\label{Section_2.2}

\begin{figure}[t]
\begin{center}
{\includegraphics[width=0.85\textwidth]{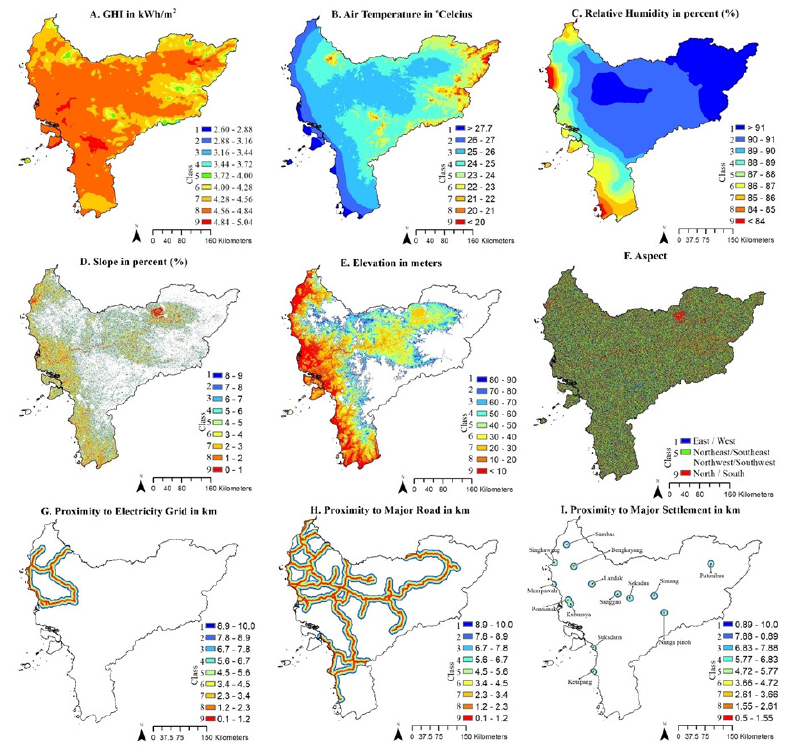}}
\caption{Reclassified layers of input criteria as shown in Table~\ref{Table_3}, with grades ranging from class 1 with low suitable value (blue color) up to the class 9 (red color) referring to maximum suitability conditions for the deployment of solar power plants. Source Maps can be consulted at our created Web-Gis within the SolarBoost project~\cite{solarboost} or by request to the author of correspondence.}
\label{Figure_5}
\end{center}
\end{figure}

To complement Table~\ref{Table_3}, it is worth mentioning that concerning the GHI and air temperature data, the World Bank Group (WBG) has already released average monthly datasets for 11 years~\cite{WBG2019}, where the monthly average deviation at WKP does not exceed the $7\%$, reason why singular (averaged) maps can be used for any of these factors, within this range of tolerance. Likewise, in order to interpolate the yearly average relative humidity data obtained as point data from NASA~\cite{NASA}, we have used the Kriging interpolation technique to generate spatial maps via the spatial analyst tools in ArcGIS 10.6.1~\cite{ESRI2004,Oliver1990}. Also, we have used the Shuttle Radar Topography Mission Digital Elevation Model (STRM-DEM) retrieved from the CGIAR Consortium for Spatial Information~\cite{CGIAR2020}, in order to prepare the ArcGIS topographical factor layers such as Elevation, Slope and Aspect maps. Thus, in Fig.~\ref{Figure_5} the spatial maps for all selected nine factors in Table~\ref{Table_3}, with their reclassified MDCA layers are shown, such that a further weighting system can be now invoked for enabling the AHP algorithm render towards a unique solution. However, as it has been mentioned at the introduction of this paper, assigning criteria factor layers and corresponding weights within an AHP algorithm is somehow a subjective process (see e.g., Table~\ref{Table_1}), as this strictly depends on the objectives of the researchers and, the extensive surveying of related literature. Thus, it results important to briefly relate why we have selected these nine factors as the most relevant criteria within the AHP algorithm, and how these are classified within a Grade-9 classification system. 

Firstly, it is worth reminding that solar collectors and PV farms are both able to utilize diffused and direct solar radiation for electricity generation, where the total amount of incoming shortwave radiation received by a horizontal surface per unit time, i.e., the Global Horizontal Irradiation, GHI in ${kWh/m^{2}}$, is the chief governing factor to locate and identify the best site to install solar farms. In simple terms, the intensity of radiation and installation area determine the magnitude of the electrical output from a solar-power plant~\cite{Yang2019,Yushchenko2018,WBG2017}, where it is known that the exploitation of solar energy resources is economically viable or profitable, specially on locations with a GHI average of $4~{kWh/m^{2}/day}$~\cite{Hernandez2015a,Hernandez2015b}. In this sense, at least from the GHI perspective, WKP is an ideal location as it attains an average of $4.58~{kWh/m^{2}/day}$ measured across a 11 years by the WBG~\cite{WBG2019}, with really scarce days reporting minimums as low as $2.6~{kWh/m^{2}/day}$ in small areas affected by short seasons of great cloudiness, but with maximums very often reaching up to $5.04~{kWh/m^{2}/day}$ in the vast majority of the province. These figures have allowed us to reclassify the daily-averaged GHI factor layer into 9 grades (Table~\ref{Table_3}), with the GIS grading classes assigned by the post-processing of the original GHI map as shown in Fig.~\ref{Figure_5}A. However, despite atmospheric factors are somehow considered within the GHI time-averaged classification above considered, and the cloudiness factor can be neglected due to the lack of solar-calendar seasons in WKP, this is not the only climatology factor of relevance for the integration of solar energy resources, as the energy conversion efficiency of these systems, in particular PV cells, strongly depend on the temperature and relative humidity conditions where they are installed. 

According to the literature, on the one hand, the efficiency of state-of-the-art PV systems increases for temperatures lower than 25~\textdegree C, but at higher temperatures, every 1~\textdegree C rise leads to a decrease in the power output of $0.4~\%$-$0.5~\%$~\cite{Doorga2019}. Hence, areas with lower average temperatures are more favorable in the context of enhancing PV system performance, and consequently must be assigned the highest grading classification, where in the specific case of WKP (Table~\ref{Table_3}), the analysis of the data provided by the WBG have revealed that, in average, the daily temperature at sunlight hours ranges from 17~\textdegree C to 28~\textdegree C, with substantial areas which might influence negatively onto the making decision process for the deployment of PV farms (see Fig.~\ref{Figure_5}B). On the other hand, the higher is the amount of relative humidity in an area the greater is the absorption of short wave solar radiation by surfaces moisture, which drops the total amount of incident solar irradiance usable by the solar panel~\cite{Abdo2013,Zhi2020}. Therefore, areas with high humidity are less prone to the exploitation of solar energy, corresponding thence to the lowest grade classes in our AHP model, with the opposite behavior for the highest classes (Table~\ref{Table_3}). This data was  obtained at 130 locations across the WKP as point data from NASA~\cite{NASA}, which is then utilized for interpolating the yearly average relative humidity by using the spatial analysis Kriging interpolation technique in ArcGIS 10.6.1~\cite{ESRI2004} as mentioned above. This technique is a powerful interpolation method based on geostatistical techniques, which allows to predict the autocorrelation relationships among the measured points whilst affording a measure of spatial accuracy on the derived map~\cite{Oliver1990}, from which we have obtained that the relative humidity in the WKP varies from $82\%$ to $91.5\%$ (see Fig.~\ref{Figure_5}C). Additionally, as mentioned above, the topographical factors have been derived from the 30-meter Shuttle Radar Topography Mission Digital Elevation Model (SRTM DEM)~\cite{CGIAR2020}, whose processed dataset is further utilized to prepare the elevation, slope, and topographical aspect maps included in Fig.~\ref{Figure_5} D-F. Therein, to reduce or even avoid the high expenditure derived from construction costs in high elevated and steep slope areas, the most favorable grade classes are given for land areas below $90~m$ in elevation, with flat or mild slopes $(<9\%)$, and with north/south facing slope (see Table~\ref{Table_3}). 

Also, the proximity to infrastructure and settlement locations is considered as one of the major factors to be included into the GIS-AHP-MDCA algorithm, as these can have a strong impact on any technical economical feasibility study of solar plants. For instance, the greater is the distance between the prospective solar plant and the existing current transmission lines, the larger will be the investment value on related infrastructure as the costs associated to transportation of specialist goods and right of way could significantly increase~\cite{Doorga2019,Majumdar2019}. Therefore, the highest-grade scale into the power-grid factor has been assigned for the areas closest to the $150~kV$ and $250~kV$ transmission lines at the WKP (Fig.~\ref{Figure_5}G), with the latter being the one connected to Malaysian power grid, and where besides invoking the grading scheme shown in Table~\ref{Table_3} with class variations within a range of $1.1~{km}$, a buffer zone of $100~m$ has been assumed in order to ensure electrical safety when working near overhead power lines~\cite{Neitzel2016}. Likewise, as the areas nearest to the major roads will avoid additional costs of transporting equipment during the construction and maintenance processes of a solar plant, a maximum radius of $10~{km}$ from road points has been considered by using the Euclidean Proximity method~\cite{Doorga2019}, where we have added a buffer of $100~m$ from either side of major roads to reduce the levels of no-natural dust sources on which the solar panels could be exposed and, even the possibility of expansion of the road by increment of the carriageways of road lanes (see Fig.~\ref{Figure_5}H). Then, the final factor to be considered is the proximity to settlements, where also a buffer of $500~m$ must be excluded from the calculations in order to reduce adverse environmental impacts on urban growth and population as suggested in~\cite{Zoghi2017}.

\subsection{Implementation of the AHP}~\label{Section_2.3}

The Analytic Hierarchy Process (AHP) is a powerful tool for MDCA which uses ratio scale factors for pairwise comparison enabling the making of a judgment or decision from the weighting of several criteria~\cite{Sanchez2013,Asakereh2017,Saaty1990,Saaty2013}. The pairwise comparison of different criteria makes AHP algorithms easy to adopt in complex GIS problems where spatial aspects can be considered by comparison of two different attributes at a time~\cite{Saaty2013,Malczewski2006, Malczewski1999}, using standard grading classification ranges as the ones shown in Table~\ref{Table_3}. Consequently, our algorithm involves a pairwise comparison matrix for nine factors with pair relative importance ranks given by the 9-point likert scale shown in Table~\ref{Table_4}, from which the relative criteria weights have been obtained by the standard AHP priority-matrix normalization method~\cite{Saaty1990}, and the AHP priority calculator introduced in~\cite{Goepel2018}. For this purpose, the priority-matrix normalization method cannot be confused with the common definition of matrix normalization in linear algebra, as in the AHP method the priority-matrix has to be normalized by dividing each assigned numerical value with the sum of values in the belonging column of the AHP priority calculator and then, the average for each row in the matrix is calculated following the strategy adopted in~\cite{Doljak2017}, which specifically applies for the MDCA of solar plants development based upon GIS-AHP algorithms.


\begin{table*}%

\caption{\label{Table_4} AHP pairwise likert grading criteria scale used for the GIS-MDCA algorithm}


\begin{center}

\begin{tabular}{ccc} \hline \hline

\textbf{Grading} & \textbf{Grade Definition by} & \textbf{Description}\\
\textbf{criteria} & \textbf{Relative Importance} & \textbf{(Ranges)}\\
\hline
1 & Equally important & Both criteria contribute equally to the objective \\
& & (within the CR)\\
2 & Mildly low & Both criteria nearly contribute equally to the objective  or, \\
& & slightly favor one criterion over another (within CR and $12.5\%$)\\
3 & Moderately low & The contribution moderately favor one criterion over another \\
& & (within $12.5\%$ and $25\%$)\\
4 & Low & The contribution has a low tendency to favor one criterion over another\\
& & (within $25\%$ and  $37.5\%$) \\
5 & Medium & The contribution has a medium tendency to favor one criterion over another\\
& & (within $37.5\%$ and  $50\%$) \\
6 & Mildly high & The contribution has slightly higher than the medium tendency to\\
& & favor one criterion over another  (within $50\%$ and $62.5\%$) \\
7 & Moderately high & The contribution  of one criterion over another is moderately high\\
& & (within $62.5\%$ and $75\%$) \\
8 & High & The contribution of one criterion over another is high\\
& & (within $75\%$ and  $87.5\%$) \\
9 & Extremely high & The contribution of one criterion over another is at the highest in the grade\\
& & (greater than $87.5\%$) \\
\hline \hline
\end{tabular}
\\
\end{center}
\end{table*}
%

Then, in order to check the consistency of the decision maker's pairwise scores, we have calculated the consistency ratio, $CR=CI/RI$, where the consistency index $(CI)$ is defined by:
%
\begin{eqnarray}\label{Eq.1}
CI=\dfrac{\lambda_{max}-n}{n-1} \, ,
\end{eqnarray}
%
%
with $\lambda_{max}$ the eigenvalue of the pairwise comparison matrix and $n$ the criteria number, where the random consistency index values, $RI$, for the $n$ values have been considered from~\cite{Saaty2013}.

Thus, to obtain meaningful results with the AHP technique within our GIS-MDCA calculations, we have ensured that for all the analyzed cases the $CR$ is less or equal to $5\%$, otherwise the pairwise comparison values are recalculated for improving the factors weighting consistency. This can be seen in Table~\ref{Table_5} where different prioritization schemes or MDCA approaches have been taken for all the nine factors within the three major considered criteria, i.e, climatology, topography, and proximity to location, being the latter the one which has a greater influence in the search for an optimal location of solar power plants, it due to the fact that the time and spatial variance of the climatology and topography factors have shown a negligible impact on their pairwise comparison with the GHI factor, as it will be shown by the sensitivity analysis shown in the following section. Consequently, the results presented in this paper are reduced to three fundamental approaches, these based upon the proximity of the solar plant to (1) the power network, (2) the road infrastructure, and (3) the community settlements, where the factor weights reported in Table~\ref{Table_5}, resulted in consistency ratios of, $CR=4.2\%$, $4.1\%$, and $4.5\%$, respectively.


\begin{table*}
\caption{\label{Table_5} Criteria and Factor weightings for the three discussed AHP approaches, each with the highest weight given either to the distance to (1) the power network, (2) the road infrastructure, or (3) the community settlements}
\begin{center}
\begin{tabular}{llllllll}
\hline \hline
\multirow{3}{*}{\textbf{Criteria}}              & \multirow{3}{*}{\textbf{AHP Factor}} & \multicolumn{3}{l}{\textbf{Factor weighting}} & \multicolumn{3}{l}{\textbf{Aggregated criteria weight}}                           \\
                                       &                             & \multicolumn{3}{l}{\textbf{by approach}}      & \multicolumn{3}{l}{\textbf{by approach}}                                          \\
                                       &                             & \textbf{1}          & \textbf{2}          & \textbf{3}          & \textbf{1}                      & \textbf{2}                      & \textbf{3}                      \\
\hline
\multirow{3}{*}{Climatology}           & GHI                         & 0.250      & 0.222      & 0.158      & \multirow{3}{*}{0.355} & \multirow{3}{*}{0.344} & \multirow{3}{*}{0.265} \\
                                       & $T$                         & 0.086      & 0.093      & 0.086      &                        &                        &                        \\
                                       & $H$                         & 0.019      & 0.029      & 0.021      &                        &                        &                        \\
\hline
\multirow{3}{*}{Topography}            & $DEM$                       & 0.026      & 0.030      & 0.027      & \multirow{3}{*}{0.114} & \multirow{3}{*}{0.150} & \multirow{3}{*}{0.128} \\
                                       & $S$                         & 0.052      & 0.071      & 0.058      &                        &                        &                        \\
                                       & $A_{z}$                   & 0.036      & 0.049      & 0.043      &                        &                        &                        \\
\hline
\multirow{3}{*}{Proximity to Location} & $G_{p}$                     & 0.272      & 0.0        & 0.0        & \multirow{3}{*}{0.531} & \multirow{3}{*}{0.506} & \multirow{3}{*}{0.607} \\
                                       & $R_{p}$                     & 0.148      & 0.351      & 0.339      &                        &                        &                        \\
                                       & $S_{p}$                     & 0.111      & 0.155      & 0.268      &                        &                        &                       
\\ \hline \hline
\end{tabular}
\end{center}
\end{table*}

%


\section{Results and Discussion}~\label{Section_3}

In a simple analysis of the GIS layers that correspond to the climatology criteria (top row in Fig.~\ref{Figure_5}), it is easy to visualize how more than $98\%$ of the WKP could be classified as a ``suitable'' area for the installation of solar power plants, if only the GHI is considered as the governing factor onto the making decisions process. Furthermore, although this decision could be refined by considering the lowest temperatures at the eastern areas of WKP, where the efficiency of solar power panels would be expected to be higher, still those areas are the ones with the highest relative humidity playing the opposite role on the overall performance of the panels. By these reasons, regardless of the AHP grading classification scheme assumed (with $CR<5\%$), if only the climatology factor is used within the GIS-AHP-MDCA algorithm, no less than $96\%$ of the WKP area could be classified as technically suitable for the deployment of solar power plants. Evidently, an area of such magnitude cannot be categorized as optimal, even when the GIS layer mapping the lands with conservation or protected status (Fig.~\ref{Figure_3} are excluded, what renders to a maximal $30\%$ of exploitable area for the deployment of solar power plants. 

The resulting $30\%$ of WKP, although already considers a strong reduction of the searchable area for the installation of solar power plants, simultaneously reducing the likelihood of finding legal barriers that could stop the investment on this technology, this simplified approach still presents significant challenges in terms of the grid-connectivity and technical deployment of solar farms. Therefore, as general rule, the topographical criteria must be considered as the secondary AHP factors, which in the case of WKP (middle row in Fig.~\ref{Figure_5}), and in general of Borneo island, these produce a nearly negligible impact onto the assessment of optimal locations of solar power plants. This is due to the fact that nearly all locations at Borneo show relatively the same Aspect conditions (relative to sun light), and the slope and elevation factors in non constrained areas (at least in WKP) show little variance on the suitability grading, i.e., with the vast majority of non or less suitable areas (classes 1-5) appearing into or in close proximity to protected or conservation areas. Therefore, the inclusion of the topography criteria lead only to a maximum optimization of the area searched from the $30\%$ of the WKP found before, towards just a $25\%-28\%$ area regardless of the AHP grading classification scheme assumed for the topography criteria.

\begin{figure}[!]
\begin{center}
{\includegraphics[width=0.99\textwidth]{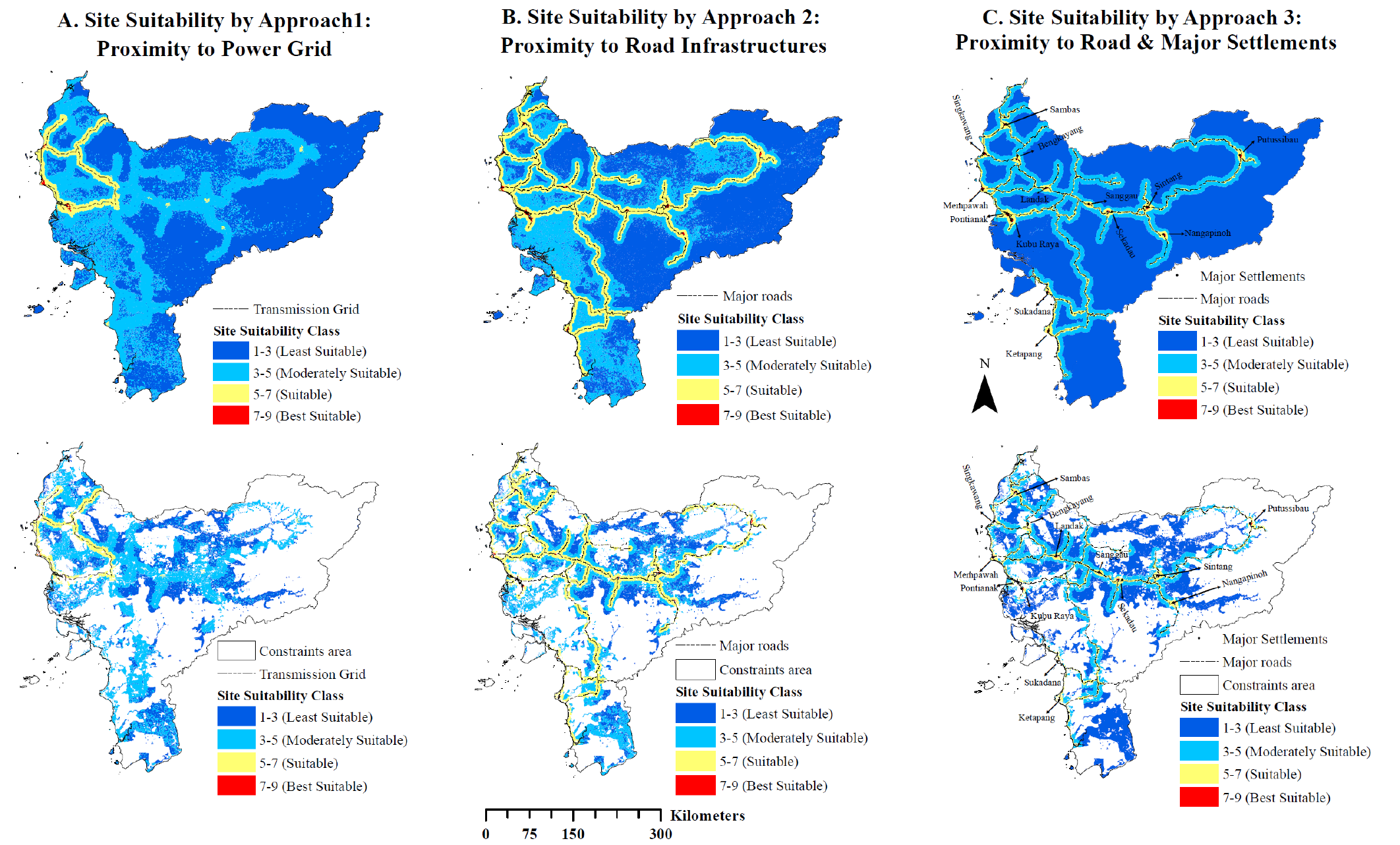}}
\caption{Spatial map showing suitability sites for large-scale deployment of solar farms at WKP before (top pane) and after (bottom pane) excluding the constraints' layer shown in Fig.~\ref{Figure_3}, under the approaches (1) proximity to the power grid (left pane), (2) proximity to road infrastructure (middle pane), and (3) proximity to roads and settlements (right pane). Class legends must be read with the highest numerical level being non-inclusive, otherwise included.}
\label{Figure_6}
\end{center}
\end{figure}

Thus, in order to provide a sensible map of optimal locations for the installations of solar power plants at WKP, which goes beyond the already comprehensive task of compiling the large set of local information for defining exclusion areas, here we focused on the relevance of the proximity factor  under the three approaches discussed above (Table~\ref{Table_5}), i.e., when the highest weighted factors are given to the shortest distances between the aimed location for the deployment of a solar plant and (1) the power transmission network, (2) the major roads, and (3) the settlements (see bottom row at Fig.~\ref{Figure_5}). Then, as the GHI across WKP has been proven to be quasi-steady and generally high along the whole year, the GHI factor is always assigned the second priority rank after the proximity factor, assuring thence a feasible amount of solar irradiation for a profitable energy production at a prospective location. Therefore, in order to simplify the analysis of the derived spatial maps from these approaches (Fig.~\ref{Figure_6}), the weighted overlay analysis of GIS layers have been grouped into four suitability categories with the ``less suitable class'' including the grades 1 to 3 from Table~\ref{Table_3} and then, the ``moderately suitable'', ``suitable'' , and ``best suitable'' classes, each grouping the grades 3 to 5, 5 to 7, and 7 to 9, respectively. Likewise, a detailed summary of the estimated areas for the optimal location of solar power plants with and without the exclusion of constrained areas is included in Table~\ref{Table_6}.

Also, it is worth noticing that depending on the technology for the conversion of solar energy and the GHI value at the region chosen, the required area for the production of a $MW$ of power can vary\cite{Gastli2010}. In this sense, the yearly electric power generation potential $(GP)$ at the WKP has been estimated from:
%
\begin{eqnarray}\label{Eq.2}
GP=SR \times CA \times SF \times \eta \times 365 \, ,
\end{eqnarray}
%
where $SR$ is the annual averaged daily GHI in $kWh/m^{2}/day$, $CA$ defines the available or suitable land area for the deployment of solar farms in $m^{2}$, $SF$ is the so-called shading factor which is an indicative measure of what fraction of the calculated areas is exploitable for PV panels (or any other solar conversion system), and $\eta$ is the solar power conversion efficiency of the system called. In this sense, we have assumed a shading factor of 0.7 based on  the maximum fraction of land that can be covered with PV solar panels with minimum shading effect~\cite{Gastli2010}, and a PV panel conversion efficiency of $\eta=16\%$ as a representative figure of the average efficiency of commercial silicon PV modules~\cite{Doorga2019}.

Thus, when the approach 1 is considered, the best suitable area for the optimal deployment of solar power plants where the highest MDCA-AHP priority factor has been given to its proximity to the power grid, followed then by the intensity of GHI, is reduced to only $230.59~{km}^{2}$, i.e, about $0.16\%$ of the WKP, which corresponds to an estimated solar energy potential for the generation of up to $43.98~{TWh/year}$. However, we have found that nearly $80\%$ of this area is actually constrained by conservation and protection laws, which limits the exploitation of solar energy to only $46.60~{km}^{2}$ of the WKP, i.e., about $8.91~{TWh/year}$, which corresponds to a daily generation capacity of approximately $43.65~{MW/km^{2}}$. This means that by utilizing approximately $18\%$ $(\sim 8.39~{km^{2}})$ of the determined area for the development of PV projects, the  national target of $366.4~{MW}$ for the WKP established by the Presidential Regulation No. 22 of 2017 (RUEN~\cite{RUEN2017}) could be achieved,  although it would demand of a considerable investment by PLN and possibly by other energy stakeholders. In fact, being the national owned company PLN the major energy stakeholder in Indonesia, we have found that their current electricity supply business plan, locally called as RUPTL~\cite{RUPTL}, lacks of sufficiently documented plans for solar energy capable to meet the RUEN targets, reason why the disclosing of suitable PV deployment areas within well documented approaches, as the ones implemented in this study, are expected to impact onto the development of local policies aimed to help the country to achieve its national targets. Moreover, given the relatively homogeneous GHI and topographical profiles across non-constrained areas at WKP (see Figs.~\ref{Figure_2}~$\&$~\ref{Figure_3}), and actually  of what is expected of any non-constrained area at national level, the estimated PV daily generation capacity of approximately $43~{MW/km^{2}}$ is not expected to change, what could allow other provinces to set up analogous policies.  

In fact, even if the less suitable areas at WKP are considered, i.e., those within the suitability classes 1-3 $(21537~{km^{2}})$, the calculated PV daily generation capacity does not vary substantially $(43.37~{MW/km^{2}})$, leading therefore to nearly the same area needed for meeting the RUEN target $(\sim 8.44~{km^{2}})$, but under the expense of possibly increasing the infrastructure costs controlled by the proximity between the solar farm and the current settlements and roads infrastructure. Thus, a much larger room for the deployment of solar power plants is foreseeable if the suitability scale considers as well the classes $5-7$, rendering to an exploitable area of about $2615.42~{km}^{2}$ ($1.78\%$ of WKP), that (in average) corresponds to areas within $3.3~{km}$ and $4.5~{km}$ of distance between the solar plant and the existing power network (see Table~\ref{Table_3}). However, the drawback of this approach is that it implies that the ``suitable'' and ``best suitable'' classifications will restrict the development of solar power plants only around the northwest of WKP, where the main power transmission network and major cities are located, deprecating  the development of further routes of commerce and communities at the central, east, and south regions of the WKP. Still, within the approach 1 these areas are somehow covered by the classes 3-5, which are considered as regions ``moderately suitable'', showing the impact of considering the proximity to the roads as a major factor within our GIS-AHP-MDCA algorithm. Nevertheless, if beyond the upgrading of the existing power grid the government decision is to extend or to increase the current capacity of the transmission network, covering then the communities above mentioned, the power grid itself which already collides with some major roads becomes irrelevant, and consequently the derived map under the approach 1 (Fig.~\ref{Figure_6}~A) could deliver an incorrect message to policy-makers. Therefore, for the development of a new power network sustained by solar power plants, a more refined approach is still required.


\begin{table*}
\caption{\label{Table_6} Full and exploitable areas for the deployment of solar power plants according to the obtained maps in Fig.~\ref{Figure_6}. Results are given in $km^{2}$ (top rows) and WKP-relative percentage (bottom rows). All numerical values under the referenced approaches are given in the mentioned units, respectively.}
\begin{center}
\begin{tabular}{lllllll}
\hline \hline
\textbf{}                                                                                & \multicolumn{2}{c}{\textbf{Approach 1 $[ km^{2} ]$}}                                                                                                                                  & \multicolumn{2}{c}{\textbf{Approach 2 $[ km^{2} ]$}}                                                                                                                                 & \multicolumn{2}{c}{\textbf{Approach 3 $[ km^{2} ]$}}                                                                                                                                 \\
\multicolumn{1}{c}{\textbf{\begin{tabular}[c]{@{}c@{}}Suitability\\ Class\end{tabular}}} & \multicolumn{1}{c}{\textbf{\begin{tabular}[c]{@{}c@{}}Full\\ 146807\end{tabular}}} & \multicolumn{1}{c}{\textbf{\begin{tabular}[c]{@{}c@{}}Exploitable\\ 48523\end{tabular}}}         & \multicolumn{1}{c}{\textbf{\begin{tabular}[c]{@{}c@{}}Full\\ 146807\end{tabular}}} & \multicolumn{1}{c}{\textbf{\begin{tabular}[c]{@{}c@{}}Exploitable\\ 48523\end{tabular}}}        & \multicolumn{1}{c}{\textbf{\begin{tabular}[c]{@{}c@{}}Full\\ 146807\end{tabular}}} & \multicolumn{1}{c}{\textbf{\begin{tabular}[c]{@{}c@{}}Exploitable\\ 48523\end{tabular}}}        \\ \hline
{ 1-3}                                                                                & 92260.59                                                                           & 21537                                                                                            & 74101.12                                                                           & 14568.91                                                                                        & 107338.7                                                                           & 27780.22                                                                                        \\
{3 - 5}                                                                                & 48985.62                                                                           & 24323.86                                                                                         & 51877.56                                                                           & 22339.82                                                                                        & 29815.7                                                                            & 15787.68                                                                                        \\
{5 - 7}                                                                                & 5330.2                                                                             & 2615.42                                                                                          & 20391.15                                                                           & 11506.28                                                                                        & 9278.1                                                                             & 4874.5                                                                                          \\
{7 - 9}                                                                                & 230.59                                                                             & 46.60                                                                                            & 437.18                                                                             & 108.58                                                                                          & 374.5                                                                              & 80.79                                                                                           \\ \hline
                                                                                         & \multicolumn{2}{c}{\textbf{Approach 1 $[ \% ]$}}                                                                                                                                      & \multicolumn{2}{c}{\textbf{Approach 2 $[ \% ]$}}                                                                                                                                     & \multicolumn{2}{c}{\textbf{Approach 3 $[ \% ]$}}                                                                                                                                     \\
\multicolumn{1}{c}{\textbf{}}                                                            & \multicolumn{1}{c}{\textbf{\begin{tabular}[c]{@{}c@{}}Full\\ 100\end{tabular}}}    & \multicolumn{1}{c}{\textbf{\begin{tabular}[c]{@{}c@{}}Exploitable \\ 33.05\end{tabular}}} & \multicolumn{1}{c}{\textbf{\begin{tabular}[c]{@{}c@{}}Full \\ 100\end{tabular}}}   & \multicolumn{1}{c}{\textbf{\begin{tabular}[c]{@{}c@{}}Exploitable\\ 33.05\end{tabular}}} & \multicolumn{1}{c}{\textbf{\begin{tabular}[c]{@{}c@{}}Full\\ 100\end{tabular}}}    & \multicolumn{1}{c}{\textbf{\begin{tabular}[c]{@{}c@{}}Exploitable\\ 33.05\end{tabular}}} \\ \hline
{1 - 3}                                                                                & 62.84                                                                              & 14.67                                                                                            & 50.47                                                                              & 9.92                                                                                            & 73.11                                                                              & 18.92                                                                                           \\
{3 - 5}                                                                                & 33.37                                                                              & 16.57                                                                                            & 35.34                                                                              & 15.22                                                                                           & 20.31                                                                              & 10.75                                                                                           \\
{5 - 7}                                                                                & 3.63                                                                               & 1.78                                                                                             & 13.89                                                                              & 7.84                                                                                            & 6.32                                                                               & 3.32                                                                                            \\
{7 - 9}                                                                                & 0.16                                                                               & 0.03                                                                                             & 0.30                                                                               & 0.07                                                                                            & 0.26                                                                               & 0.06                                                                                            \\ \hline \hline
\end{tabular}
\end{center}
\end{table*}

Based in the observations above, and by noticing that the electricity grid is also at the proximity to major roads connecting the cities of Sambas, Singkawang, Menpawah, Pontianak, Kuburaya, Landak, and Bengkayang (see bottom row of Fig.~\ref{Figure_5}), it is possible to center the proximity to location criteria to only the factors of Road Proximity, $R_{p}$, and Settlement Proximity, $S_{p}$ as shown in the approaches 2 and 3 of Table~\ref{Table_5}. Within these strategies, is then important to re-assess the weighting criteria for each one of these factors, such that comparison between the different schemes is viable, by ensuring that the same or nearly the same consistency ratio is obtained as mentioned in section~\ref{Section_2.3}. Thus, in the approach 2, the $R_{p}$ is given the highest priority followed by the GHI, as this will be the most natural approach to take, keeping similar ratios for the other factors as established in the approach 1. In this way, we have found that the ``best suitable'' unconstrained sites for the deployment of solar power plants can be in an area as large as $437.18~{km}^{2}$ ($0.3\%$ of the WKP), which is a bit less than twice the unconstrained area obtained within the approach 1. However, by excluding the constrained regions we have found that the total exploitable area ($108.58~{km}^{2}$) is actually of about 2.33 times greater than the one obtained with the approach 1, which is an excellent result as this now covers the eastern and southern areas of the WKP that were not deemed as ``best suitable'' in the approach 1. 

Nevertheless, a couple of issues arise with the interpretation of the GIS map derived by the approach 2, as although the settlement locations in Fig.~\ref{Figure_5}I could be easily identified to be at the center of the denser ``best suitable'' areas (large red dots), with these and the classified ``suitable'' areas within approximately $20~{km}$ from the main road network, there are also ``moderately suitable'' regions which are much more than $20~{km}$ far away of the settlements, the current roads and, the power grid infrastructures, a fact which will considerably rise the cost of development of a solar power plant. Thus, despite these areas could be considered as good locations for new settlements and further expansion of the WKP communities, if a better classification of suitable and best suitable sites for the deployment of solar farms is to be given to policy makers, investors, and other stakeholders, an additional approach where greater weight is to be given to the proximity to existing settlements needs to be pursued.  

Consequently, in the approach 3 we have increased the relative weighting of the $S_{p}$ factor, at the expense of reducing the importance of the GHI due to the established isomorphism of this condition. This resulted in a most refined suitability map, where not only the ``best suitable'' exploitable area which in average covers the grade classes $7-9$ in Table~\ref{Table_3}, and the corresponding ``suitable'' ones for the classes 5-7, both approximately double the solar exploitation areas obtained within the approach 1, thereby including the largely dependent communities on diesel-powered electricity such as Sintang, Sanggau, Nanga Pinoh, Sukadana, and Ketapang, but also present a clearer and most systematic distinction on what accordingly with Tables~\ref{Table_3}~\&~\ref{Table_6} can be considered as the regions of WKP which are moderately or least suitable for the deployment of solar power plants.

The total annual electric power generation capacity at the WKP has been also calculated before and after excluding the constraint conditions for each one of the three approaches above considered (see Table~\ref{Table_7}). This shows the large effect that the consideration of protected and conservation areas imply on measuring the energy potential at the diverse regions of Borneo Island, and how the classification of priority criteria can render to largely different predictions. Still, we have demonstrated the enormous potential of Borneo island in what concerns to solar energy production, as a proper planning of solar power plants just at the WKP could be sufficient to meet the 2030 clean energy plans of Indonesia~\cite{RUPTL,PLN_RUPTL2019}. Moreover, regardless on the plans of the Indonesian government, either by expanding the current power transmission network to supply and connect the cities or settlements largely dependent on fossil energy production, or to upgrade the existing power transmission network for achieving a better and more competitive interconnection with the Malaysian power grid at the North-west of the WKP, we have found that a common factor within the three adopted approaches is that the optimal site for the deployment of solar power plats resulted in locations proximate with the city of Mempawah, therefore, also close to the cities of Pontianak and Singkawang, with an average exploitable area of 3019 hectares and annual solar energy production capacity of $6.63~{TWh}$. Likewise, the nearby of the city of Putussibau at the eastern side of WKP is also highlighted, as the approaches 2 and 3 have allowed us to identify about $1689~Ha$ (average) as the best suitable locations for solar power plants, which could be better positioned than the "best suitable" areas closer to the cities of Sintang, Sanggau, Sekadau, Nanga Pinoh, Sukadana, and Ketapang, given its good road connection and proximity with the 
major AH150 road at Malaysia.  


\begin{table*}
\caption{\label{Table_7} Projected annual PV energy generation potential at the full and exploitable areas of WKP accordingly with Eq.~\ref{Eq.2} and the results obtained in Table~\ref{Table_6} for the three approaches adopted in this study (see Fig.~\ref{Figure_6}). All numerical values under the referenced approaches are given in $TWh/year$.}
\begin{center}
\begin{tabular}{lllllll}
\hline
\textbf{}                                                                                & \multicolumn{2}{c}{\textbf{Approach 1 $[ TWh/year ]$}}                                                                                                                            & \multicolumn{2}{c}{\textbf{Approach 2 $[ TWh/year ]$}}                                                                                                                            & \multicolumn{2}{c}{\textbf{Approach 3 $[ TWh/year ]$}}                                                                                                                            \\
\multicolumn{1}{c}{\textbf{\begin{tabular}[c]{@{}c@{}}Suitability\\ Class\end{tabular}}} & \multicolumn{1}{c}{\textbf{\begin{tabular}[c]{@{}c@{}}Full\\ 27530.18\end{tabular}}} & \multicolumn{1}{c}{\textbf{\begin{tabular}[c]{@{}c@{}}Exploitable\\ 9261.67\end{tabular}}} & \multicolumn{1}{c}{\textbf{\begin{tabular}[c]{@{}c@{}}Full\\ 27534.56\end{tabular}}} & \multicolumn{1}{c}{\textbf{\begin{tabular}[c]{@{}c@{}}Exploitable\\ 9247.31\end{tabular}}} & \multicolumn{1}{c}{\textbf{\begin{tabular}[c]{@{}c@{}}Full\\ 27541.30\end{tabular}}} & \multicolumn{1}{c}{\textbf{\begin{tabular}[c]{@{}c@{}}Exploitable\\ 9253.48\end{tabular}}} \\ \hline
{[}1-3{]}                                                                                & 17106.25                                                                             & 4090.97                                                                                    & 13648.14                                                                             & 2758.76                                                                                    & 20022.66                                                                             & 5287.95                                                                                    \\
{[}3-5{]}                                                                                & 9365.94                                                                              & 4663.51                                                                                    & 9909.41                                                                              & 4271.06                                                                                    & 5671.83                                                                              & 3015.33                                                                                    \\
{[}5-7{]}                                                                                & 1014.01                                                                              & 498.28                                                                                     & 3892.90                                                                              & 2196.54                                                                                    & 1774.81                                                                              & 934.59                                                                                     \\
{[}7-9{]}                                                                                & 43.98                                                                                & 8.91                                                                                       & 84.11                                                                                & 20.96                                                                                      & 72.00                                                                                & 15.62                                                                                      \\ \hline \hline
\end{tabular}
\end{center}
\end{table*}
%

Finally, for concluding the MDCA process, it is important to conduit a sensitivity analysis of the GIS-AHP outcomes, such that a policy or decision maker can give suitable references onto the influence of each input criteria and associated priority factors in the attained suitability maps. Generally, AHP results are highly sensitive to the weighting criteria and prioritization factors which depend directly on the specific objectives of the research involved or the aims of particular stakeholders~\cite{AlGarni2017,Doorga2019}, which therefore implies that there is not a unique method for defining a sensitivity analysis which could not be seen, somehow, biased towards one or another factor. However, a systematic sensitivity analysis can be performed by excluding individual input criteria in the AHP analysis, such that the influence of this particular criteria can be quantified from the GIS-AHP suitability maps obtained, as long as the chief criteria is maintained across the entire study, i.e., the GHI data layer in our case. Thus, as by excluding a specific input criteria into the AHP algorithm, the sensitivity analysis demands to keep the relative weightings of the other criteria unchanged, then, due to possible changes in the pairwise comparison system, a small fluctuation $\Delta S$ in the AHP weightings appears~\cite{Kumar2016,Doorga2019}, which can be accounted by
%
\begin{eqnarray}\label{Eq.3}
\Delta S_{i,j} = \dfrac{S_{i,j} - S_{j}}{S_{j}} \times 100\% \, ,
\end{eqnarray}
%
%
where $\Delta S_{i}^{j}$ is the percentage change in the \textit{j-th} site suitability class due to the exclusion of the assessed \textit{i-th} input criteria in Table~\ref{Table_3}, with $i\neq j$, and $S_{i,j}$ and $S_{j}$ the corresponding suitability class areas (see Fig.~\ref{Figure_6}) with and without the exclusion of the \textit{i-th} input criteria. 

Thus, as all criteria with exemption of the GHI can be omitted, the remaining 8 AHP factors are individually assessed, with the resulting sensitivity values being shown in Table~\ref{Table_8}. The results indicate both positive and negative changes in the percentage areas for the specific suitability classes, i.e., increments or decrements of the calculated areas  due to the exclusion of individual input criteria, respectively. In particular, it is to be noticed that starting from the approach 1, and by excluding the proximity to the power grid factor, $G_{p}$, the area originally classified as ``best suitable'' increases in only a $14.66\%$. This is due to the now larger influence of the other two proximity factors, road infrastructure and settlements' location, which as a matter of fact have a much lower impact in the GIS area calculations. Consequently, a large increment in the other classification areas, i.e., those called as ``moderately suitable'' and ``suitable'' would be seen as a result of the larger areas covered by the topography and climatology factors, which in average tend to benefit the western side of the WKP (See Table~\ref{Table_5} \& Figure~\ref{Figure_5}). This explains then, why by excluding the proximity to the road factor, $R_{p}$, both extremes of the GIS-AHP map classification, i.e., the least and best suitable areas, both result reduced in about $50\%$ the originally calculated areas (see Table~\ref{Table_6}), as the moderately suitable area will be now predominately defined by the $G_{p}$ and $S_{p}$ factors. This makes the areas closer to the settlements and the current power transmission network to be more favorable for the deployment of solar farms, but at the expense of possible considerations in the expansion of the power network to the large number of settlements currently dependent of power-diesel generators, as previously discussed. Moreover, if the $S_{p}$ factor is the one to be excluded, then finding an optimal location for the development of a solar farm could be more biased, as regardless of the inherent social and economic development implications that this decision might imply, the  ``best suitable'' area would increase in more than $400\%$ the original found area, which is not precisely a better result from the technical and investor points of view, as as farther is the distance between the power network at the transmission level to the settlements, as larger will need to be the power distribution-level network that will need to be created. Consequently, although the sensitivity analysis presented in table~\ref{Table_8} allows to quantify in a simple manner the influence of the different GIS-AHP weighted factors into the present MDCA, it also reveals why the three approaches presented across this paper focus mainly on the strong dependence of the proximity criteria, as generally the GIS-AHP-MDCA implemented at the WKP is less sensitive to changes onto the climatology and topography factors.     


\begin{table*}
\caption{\label{Table_8} Percentage sensitivity factor $\Delta S_{i}^{j}$ as described by Eq.~\ref{Eq.3}, with the $i-$input criteria defined as in Tables~\ref{Table_3}~\&~\ref{Table_5}.}
\begin{center}
\begin{tabular}{lllll}
\hline \hline
\textit{\textbf{}}                                                                                     & \multicolumn{4}{c}{\textbf{$\Delta S_{i}^{j}$~$[\%]$}}                                                                                                                                                                                                                                                                                                                         \\
\multicolumn{1}{c}{\textit{\textbf{\begin{tabular}[c]{@{}c@{}}Excluded \\ criteria - i\end{tabular}}}} & \multicolumn{1}{c}{\textbf{\begin{tabular}[c]{@{}c@{}}$j =$\\ Least Suitable\end{tabular}}} & \multicolumn{1}{c}{\textbf{\begin{tabular}[c]{@{}c@{}}$j =$\\ Moderately Suitable\end{tabular}}} & \multicolumn{1}{c}{\textbf{\begin{tabular}[c]{@{}c@{}}$j =$\\ Suitable\end{tabular}}} & \multicolumn{1}{c}{\textbf{\begin{tabular}[c]{@{}c@{}}$j =$\\ Best Suitable\end{tabular}}} \\ \hline
$T$                                                                                                    & 6.60                                                                                        & -12.60                                                                                           & -0.95                                                                                 & 234.62                                                                                     \\
$H$                                                                                                    & 3.53                                                                                        & -6.83                                                                                            & 2.05                                                                                  & 29.88                                                                                      \\
$DEM$                                                                                                  & 10.01                                                                                       & -19.08                                                                                           & -2.18                                                                                 & 10.53                                                                                      \\
$S$                                                                                                    & 5.80                                                                                        & -11.79                                                                                           & 2.54                                                                                  & 50.82                                                                                      \\
$A_{z}$                                                                                              & 8.99                                                                                        & -17.57                                                                                           & -2.61                                                                                 & 101.40                                                                                     \\
$G_{p}$                                                                                                & -95.94                                                                                      & 148.63                                                                                           & 315.50                                                                                & 14.66                                                                                      \\
$R_{p}$                                                                                                & -54.73                                                                                      & 105.94                                                                                           & -7.87                                                                                 & -52.86                                                                                     \\
$S_{p}$                                                                                                & -42.38                                                                                      & 77.27                                                                                            & 11.18                                                                                 & 471.64                                                                                    
\\ \hline \hline
\end{tabular}
\end{center}
\end{table*}
%


\section{Conclusions}\label{Section_4}

In order to meet the Indonesian goal of increasing the share of renewables from $12.4\%$ in 2020 to $28.6\%$ by 2030~\cite{Irena2017,RUEN2017}, the development of sustainable energy plans with a large intake of renewables is currently the focus of major governmental and private energy investors ~\cite{RUPTL,Maulidia2019}. However, the lack of a unique coordination platform for planning resources and data driven MDCA-tools aimed to coordinate regional information by diverse stakeholders, facilitating thence the exploitation of renewables, has led the Indonesian government to focus their efforts into the exploitation of geothermal reserves at the provinces of Java and Sumatra~\cite{Bloom2019}, as these are not only well located but also not constrained by enforcement of conservation or protection laws. In this sense, in this paper we have presented the major results derived from the UK-Indonesia Institutional Links project ``SolarBoost'' funded by the British Council, where a unique GIS-AHP-MDCA platform has been created for disclosing the main physical layers that can infer in the determination of the optimal location of solar power plants at the West Kalimantan Province (WKP), being this the second largest but most populated region of the Indonesian Borneo. It is worth mentioning that WKP has been classified of special interest for this project, as the Indonesian government has recognized WKP as the region with the greatest potential for solar energy generation ~\cite{RUEN2017}, with a projected electricity consumption of $4,309~{GWh}$ for 2028 exceeding in at least $25\%$ the maximum capacity of an already overburden power grid~\cite{PLN_RUPTL2019}.

Thus, thanks to the collaboration with a large set of stakeholders (see Refs.~\cite{PLN,ESDM,BAPPEDA,MENLHK,PERTANIAN,PUPR,BMKG}), here we report a unified GIS map (Fig.~\ref{Figure_3}), where diverse sources of digital and printed information, originally retrieved in different formats (See Table~\ref{Table_2}~\& Fig.~\ref{Figure_4}), is compiled for demonstrating that although $96\%$ of the WKP can be classified as highly suitable for the deployment of solar power plants, this if only the Global Horizontal Irradiation (GHI) is considered as the decision criteria, at least $66\%$ of the WKP could not be exploited as this would involve areas subjected the existence of land conservation and protection laws. In particular, we have found that although protected and conservation forests cover up to $25.5\%$ of WKP, an additional $30.2\%$ of forestry must be also considered to include the protected production rights that the the lumber industry has (see Fig.~\ref{Figure_4}). Likewise, GIS maps for other protected and conservation areas have been retrieved, including detailed spatial information for rice fields $(2.1\%)$, peatlands $(4.2\%)$, community and wildlife forests $(9.1\%)$, water bodies $(2.5\%)$, and Orang-Utan habitats $(19.9\%)$. 

On the other hand, taking benefit of the geographic location of WKP (Fig.~\ref{Figure_1}, which is crossed by the equatorial line, single averaged maps have been created from the set of 11-years monthly derived GHI and air temperature data provided by the World Bank Group~\cite{WBG2019}, within a tolerance factor of $7\%$. Likewise, the yearly average relative humidity map has been obtained as point data from NASA~\cite{NASA}, where we have used the Kriging interpolation technique to generate spatial maps via the spatial analyst tools in ArcGIS 10.6.1~\cite{ESRI2004,Oliver1990}. Then, topography factors such as elevation, slope, and aspect have been obtained with the STRM digital elevation model from CGIAR~\cite{CGIAR2020}. Finally, in Fig.~\ref{Figure_5} we show the resulting spatial maps with reclassified MDCA layers for the AHP weighting processing of all the nine criteria factors included in Table~\ref{Table_3}, where a comprehensive analysis on each one of these factors and their relevance on the determination of optimal locations for the deployment of solar power plants have been included. Within these factors, we have demonstrated why after considering the exclusion of constrained areas, the proximity to infrastructure and settlement locations play a major role into the AHP-MDCA algorithm for the analysis of the GIS layers.    

For this purpose, within the 9-point likert scale shown in Table~\ref{Table_4}, we have developed an AHP-MDCA algorithm which involves a pairwise comparison of the nine GIS factors aforementioned via the priority-matrix normalization method~\cite{Doljak2017} and, the use of the AHP online priority calculator introduced in~\cite{Goepel2018}. Three different approaches within an acceptable consistency ratio, $CR<5\%$, 
have been considered for providing sensible maps for the optimal location of solar power plants at WKP, beyond the exclusion  of conservation and protected areas reported above (see Table~\ref{Table_5} and Fig.~\ref{Figure_6}). These approaches focus the on weighting factors accounting for the proximity of the intended solar plant and the existing (1) power transmission network, $G_{p}$, (2) roads infrastructure, $R_{p}$, and (3) settlements, $S_{p}$ as these factors are known to play a major role on the economic costing and overall planning of a solar plant.

Thus, by considering the power generation potential model in Eq.~\ref{Eq.2}, and by assuming a solar power conversion factor of $16\%$, we have determined that when the greatest priority is given to $G_{p}$ (approach 1), the search for an optimal area for the deployment of solar power plants with the best suitable conditions is reduced to a $0.16\%$ of the WKP area, which after exclusion of constrained areas is further reduced to just $46.6~{km}^{2}$, or $0.03\%$ of WKP, attaining an area nearly 60 times greater if the moderately suitable conditions are accepted (see Table~\ref{Table_6}). However, we have demonstrated that our finding for the best suitable areas classification, results sufficient to meet solar capacity target of the Indonesian government~\cite{Maulidia2019}, if only a relatively small $2~{km}^{2}$ PV solar power plant is developed per year. Nevertheless, within the MDCA we have argued that the ``best suitable'' and ``suitable'' classifications within this approach, can both present a serious drawback for policy makers if it is not contrasted with other relevant approaches, as it tends to contain the development of solar power plants around the northwest of WKP. Therefore, in the approach 2, we have given a greater weighting to $R_{p}$, avoiding the double spatial weighting that is given when the $G_{p}$ factor is also considered, as the major power network at WKP is at the proximity of some of its major roads. In this way, the overall dimensions of the best suitable area is now about 2.3 times greater than the obtained one  in the approach 1, and considers now regions of potential interest at the eastern and southern areas of WKP. However, still the exploitable area for the regions classified as ``moderately suitable'' (after excluding constraints), result to be the largest of all the classes within the approach 1 and approach 2, showing a lower correlation with the semi-qualitative meanings given to the four suitability classes within our MDCA when the full WKP area is considered (see Table~\ref{Table_6}). In other words, when the constrained areas are not included, the largest area can be always identified within the suitability classes 1-3 and, as the suitability level increases the area of scope is reduced towards most optimal regions. This allows a convenient correlation between the criteria grades (Table~\ref{Table_3}) and the suitability classes (Table~\ref{Table_6}) for policy makers and stakeholders, which unfortunately cannot be guaranteed when the constraint areas are excluded. In this sense, we have found that this convenient representation of suitability levels for policy makers and potential investors in the solar power market at WKP, can be ensured by a proper factor weighting of the distance between the prospective solar power plant and current settlements with the climatology and topography factors (approach 3), ensuring the $CR\leq 5\%$ when lowering the weighting of those factors less sensible to the GIS-AHP-MDCA (Table~\ref{Table_8}), a process from which we have determined that out of the maximum exploitable area ($33.05\%$ of WKP, regardless the AHP approach), its $18.92\%$ can be considered as ``least suitable'', $10.76\%$ as `` moderately suitable'', $3.32\%$ as ``suitable'', and $0.06\%$ ($80.79~{km}^{2}$) as ``best suitable'', the latter corresponding to a remarkable year-projected PV generation potential of up to $15.62~{TWh}$, which could be sufficient to meet the 2030 clean energy plans of Indonesia~\cite{RUPTL,PLN_RUPTL2019}. 

Thus, although the three approaches considered in this paper have render to the estimation of highly suitable areas for the deployment of solar power plants at WKP, where a stakeholder could give a conscious preference to one or another factor, we argue that from our single perspective, our third approach allows a better or clearer inclusion of settlements with electricity networks largely dependent on diesel generators, such as the ones at the towns of Sintang, Sanggau, Nanga Pinoh, Sukadana, and Ketapang, presenting as well a clearer and systematic distinction onto the diverse suitability levels when the region is heavily subject to non-physics or legally based constraints. This has shown the large effect that the consideration of protected and conservation areas imply on measuring the energy potential at WKP, and consequently on the diverse regions of Borneo Island.


\section*{Acknowledgement}

This research used the ALICE High Performance Computing Facility at the University of Leicester. This work was supported by the Institutional Links grant, ID 413871894, under the Newton Fund Indonesia partnership. The grant is funded by the UK Department for Business, Energy and Industrial Strategy and the Indonesian Ministry of Research, Technology, Higher Education and delivered by the British Council. For further information, please visit \url{www.newtonfund.ac.uk}. HSR (UK) and AS (IND), principal investigators of this project (SolarBoost), express their special thanks to the set of stakeholders and project partners (see Refs.~\cite{PLN,ESDM,BAPPEDA,MENLHK,PERTANIAN,PUPR,BMKG}) for the valuable discussions and sharing of data.

%
%

\bibliographystyle{elsarticle-num}
\bibliography{References_Ruiz_Group}

\end{document}